\documentclass[prl,aps,twocolumn,showpacs,nofootinbib,nobibnotes,superscriptaddress]{revtex4}

\usepackage{epsfig}
\usepackage{amssymb}

\begin{document}


\title{Polarized Cosmological Gravitational Waves  from Primordial
Helical Turbulence}

\author{Tina Kahniashvili}
\affiliation{Department of Physics, Kansas State University,
116 Cardwell Hall, Manhattan, KS 66506, USA}
\affiliation{Center for Plasma Astrophysics, Abastumani Astrophysical
Observatory, 2A Kazbegi Ave, GE-0160 Tbilisi, Georgia}
\author{Grigol Gogoberidze}
\affiliation{Center for Plasma Astrophysics, Abastumani Astrophysical
Observatory, 2A Kazbegi Ave, GE-0160 Tbilisi, Georgia}
\author{Bharat Ratra}
\affiliation{Department of Physics, Kansas State University,
116 Cardwell Hall, Manhattan, KS 66506, USA}

\begin{abstract}
We show that helical turbulence produced during  a first-order phase
transition generates circularly polarized cosmological
gravitational waves (GWs).
The characteristic frequency of these GWs
for an extreme case of the phase transition model is
around $10^{-3}$ --- $10^{-2}$ Hz with an energy
density parameter as high as $10^{-12}$ --- $10^{-11}$.
The possibility of detection
 is briefly discussed. \end{abstract}


\maketitle

Since cosmological GWs propagate without significant
interaction after they are produced, once detected they should provide
 a powerful tool for studying the early
Universe at the time  of GW generation
\cite{tasi}.
 Various mechanisms  for cosmological GW generation have been studied,
including: quantum fluctuations
during inflation \cite{inflation};
  bubble wall motion
and collisions during phase transitions \cite{kos};
 cosmological magnetic fields \cite{magnet,kmk02};
 and plasma turbulence
\cite{kmk02,dolgov,nic}.

In this letter we focus on polarization of
cosmological GWs generated by helical
 stochastic turbulent motions \cite{helicity,helicity2,kr05}.
 We find that helical
turbulence generates circularly polarized stochastic GWs
and we compute the polarization degree.
 The  formalism we use is
general and can be applied to study the generation of stochastic
GWs  by
any helical vector field (e.g.,
helical magnetic fields \cite{cdk04,kr05}). Primordial polarized  GWs
might be  generated from quantum fluctuations accounting for
 the gravitational
Chern-Simons term \cite{rodriguez}.

GWs are sourced by the transverse and traceless part of the
stress-energy tensor $T_{\mu \nu}$ \cite{mtw73}.
In the case at hand $T_{\mu \nu}$ describes  a turbulent cosmological fluid after a
 phase transition \cite{kos,kmk02,dolgov}. For spatial indices $i \neq
j$, $T_{ij}({\bf x}) = (p+\rho) u_i({\bf x}) u_j({\bf x})$,  where $p$ and
$\rho$ are the fluid pressure and  energy density and
 ${\mathbf u}({\mathbf x})$
 is the fluid velocity.
 The fluid enthalpy density $p+\rho$ is taken  to be
constant throughout space.
The transverse and traceless part of
$T_{ij}$ in Fourier space is  $ \Pi_{ij}({\mathbf
{k}})=[ P_{il}({\mathbf{\hat k}})P_{jm}({\mathbf{\hat k}})
-\frac{1}{2} P_{ij}({\mathbf{\hat k}}) P_{lm}({\mathbf{\hat k}})
]T_{lm}({\mathbf{ k}})$, where
$P_{ij}({\mathbf{\hat k}}) = \delta_{ij}-{\hat k}_i {\hat k}_j$
with  ${\hat k}_i = k_i/k$.
Consistent with observations we
have assumed flat space sections.

To model the turbulence we assume  that in the early Universe at time
$t_{\rm{in}}$ (at a phase transition)
liberated vacuum energy $\rho_{\rm{vac}}$
is converted into  (turbulent) kinetic
energy of the cosmological plasma  with an efficiency
$\kappa$ over a time scale $\tau_{\rm{stir}}$
 on a  characteristic source length scale $L_S$
 \cite{kos}.
 After  generation, the turbulence kinetic energy  cascades from
larger to smaller scales. The cascade stops at a damping scale,
$L_D$, where the turbulence  energy is removed by
dissipation.
As usual, we  assume
 that the turbulence  is produced in a time much less
than the Hubble time,  $\tau_{\rm{stir}} \ll 1/H_{\rm{in}}$ --- here
$H_{\rm{in}}$ is the Hubble parameter at $t_{\rm{in}}$ ---
 \cite{kmk02,dolgov}, and therefore we
ignore the expansion  of the Universe  when
 studying the  generation of GWs. In this case the GW equation of motion,
 in wave number space, is \cite{mtw73}
\begin{equation}
{\ddot h}_{ij}({\mathbf k}, t) + k^2 h_{ij}({\mathbf k}, t) = 16\pi
G \Pi_{ij} ({\mathbf k}, t). \label{h_evolution}
\end{equation}
Here $G$ is the Newtonian gravitational constant,
 and $h_{ij}({\mathbf k})=\int d^3\!x \,
   e^{i{\mathbf k}\cdot {\mathbf x}} h_{ij}({\mathbf x})$
and $h_{ij}({\mathbf x})=\int d^3\!k \,
   e^{-i{\mathbf k}\cdot {\mathbf x}} h_{ij}({\mathbf k})/(2\pi)^3$
is the Fourier transform pair of the tensor metric
 perturbation which is defined
$h_{ij}=\delta g_{ij}$   ($h_{ii}=0$ and $h_{ij}{\hat k}^j=0$). We
use natural units $\hbar = 1 = c$, physical/proper wave numbers
(not comoving ones),
 and an overdot denotes a
 derivative with respect to
time $t$.

Stochastic turbulent fluctuations
generate stochastic  GWs. Gaussian-distributed GWs may be characterized by
 the wave number-space two-point  function
\begin{eqnarray}
\langle h^{\star}_{ij}({\mathbf k},t) h_{lm} ({\mathbf k'},t)\rangle
~~~~~~~~~~~~~~~~~~~~~~~~~~~~~~~~~~~~~~~~~~
\nonumber\\
=(2\pi)^3 \delta^{(3)}({\bf k}-{\bf k'}) \left[ {\mathcal
M}_{ijlm} H(k,t)
+ i{\mathcal A}_{ijlm} {\mathcal H} (k,t) \right].~
\label{gw1}
\end{eqnarray}
Here $H ({k}, t)$ and ${\mathcal H}({k}, t)$ characterize
the GW amplitude and polarization,
 $4 {\mathcal M}_{ijlm} ({\mathbf{\hat k}}) \equiv
P_{il}P_{jm}+P_{im}P_{jl}-P_{ij}P_{lm}$, and
 $8 {\mathcal A}_{ijlm}({\mathbf{\hat k}})
\equiv {\hat {\bf k}}_q (P_{jm} \epsilon_{ilq} + P_{il}
\epsilon_{jmq} + P_{im} \epsilon_{jlq} + P_{jl} \epsilon_{imq})$ are
tensors, and $\epsilon_{ijl}$ is the fully antisymmetric symbol.
 Choosing the coordinate system so that unit vector
 ${\hat {\mathbf e}}_3$ points in the
GW propagation  direction, using the usual circular polarization basis
tensors  $e^{\pm}_{ij} = -({\bf e}_1
\pm i{\bf e}_2)_i \times ({\bf e}_1 \pm i {\bf e}_2)_j/\sqrt{2}$
 \cite{mtw73},  and
defining two states
$h^+$ and $h^-$ corresponding  to right- and left-handed
 circularly  polarized
GWs,  we have $h_{ij}=h^+e^+_{ij} + h^-
e^-_{ij}$. The GW degree of circular polarization  is given by \cite{meszaros}
\begin{equation}
{\mathcal P}(k) =  \frac {\langle h^{+ \star}({\mathbf k})
h^{+}({\mathbf k'}) -
 h^{- \star}({\mathbf k}) h^{-}({\mathbf k'}) \rangle}
{\langle h^{+ \star}({\mathbf k}) h^{+}({\mathbf k'}) +
 h^{- \star}({\mathbf k}) h^{-}({\mathbf k'}) \rangle}
=\frac{{\mathcal H}(k)}{H(k)}.~ \label{degree}
\end{equation}

 Both  $H(k,t)$ and ${\mathcal H}(k,t)$
 are  obtained by solving Eq.~(\ref{h_evolution}), and are
related to $\Pi_{ij}({\mathbf k},t)$. For instance, an
axisymmetric stochastic vector source (non-helical turbulent
motion or any other non-helical vector field) induces unpolarized GWs with $
|h^+({\bf k}, t)|  = |h^-({\bf k}, t)|$
\cite{magnet,kmk02,dolgov};
 the presence of
 a helical source
 alters this situation.

To compute the induced GW power spectrum one must have the
 source two-point function
$\langle \Pi_{ij}^\star ({\mathbf k}, t) \Pi_{lm}
({\mathbf k^\prime},t^\prime )\rangle $.  This is determined
by the fluid velocity two-point  function.
For stationary,
isotropic and homogeneous flow the velocity
two-point  function is \cite{pvw02,kr05}
\begin{equation}
\langle u^\star_i({\mathbf k})u_j({\mathbf k'})\rangle ={(2\pi)^3}
\delta^{(3)}({\mathbf k}-{\mathbf k'}) [P_{ij} P_S(k) + i \epsilon_{ijl}
\hat{k}_l P_H(k)]. \label{spectrum}
\end{equation}
Here $P_S(k)$ and $P_H(k)$ are
the symmetric (related to the
kinetic energy density per unit enthalpy of the fluid)
 and helical (related to the  average
kinetic helicity $\langle \mathbf{u}
 \cdot (\nabla \times \mathbf{u}) \rangle$) parts of the
 velocity
power spectrum \cite{pvw02,kr05}. Causality requires $P_S(k) \geq
|P_H(k)|$; see  p.~161 of Ref.~\cite{L}.

However, in the case of interest here,  the source of turbulence
 acts for only a short time $\tau$, possibly not  exceeding
the large-scale eddy turnover time $\tau_S$ (corresponding  to
 length scale  $L_S$) --- for self-consistency, however,
 we assume that the source is active
 over a time $\tau =  \mbox{max}( \tau_{\rm{stir}}, \tau_S)$
 \cite{kmk02} --- resulting in a time-dependent velocity spectrum.
To model the development  of  helical
turbulence during the time interval
 $(t_{\rm{in}}, t_{\rm{fi}}=t_{\rm{in}}+\tau)$
we make several simplifying assumptions:

(a) Turbulent fluid kinetic energy is present on all
scales in the inertial
 range  $k_S<k<k_D$. Here $k_S=2\pi/L_S$ and $k_D=2\pi/L_D$.
We also assume that the energy is injected into the turbulence
continuously over a time $\tau$,
rather than as an instantaneous impulse \cite{kmk02,dolgov}.

(b) Unequal time correlations are  modeled as \cite{kmk02}
\begin{eqnarray}
\langle u^\star_i({\mathbf k}, t) u_j({\mathbf k'}, t')\rangle
&=&{(2\pi)^3} \delta^{(3)}({\mathbf k}-{\mathbf k'}) [P_{ij} F_S(k,
t-t^\prime)
\nonumber\\
&&~~~~~~~~ + i
\epsilon_{ijl} \hat{k}_l F_H(k, t-t^\prime)], \label{spectrumtime}
\end{eqnarray}
where  the $t-t^\prime $
 dependence of the  functions $F_S$ and $F_H$
 reflects the assumption of time translation invariance.
Since energy is injected continuously, at  $t=t^\prime\in (t_{\rm{in}},
t_{\rm{fi}})$,  $F_{S}(k,0)=P_{S}(k)$ and $ F_{H}(k,0)=P_{H}(k)$.

(c) The decay of non-helical   turbulence is determined
 by a monotonically decreasing function $D_1(t)$
and $ F_S(k,t) = P_S(k) D_1(t)$, p.~259 of Ref.~\cite{H}. Extending
this assumption to the  helical turbulence case we also model
$F_{H}(k,t)=P_{H}(k) D_{2}(t)$, where
 $D_{2}(t)$ is another monotonically decreasing function.
 Since in the considered model most of the power
is
 in the inertial range,
 for simplicity we discard power outside the inertial range by truncating
 $P_S$ and $P_H$ at  $k<k_S$ and $k>k_D$.

(d) We model the power spectra by  power laws,
$P_{S}(k) \propto k^{n_{S}}$ and $P_{H}(k) \propto k^{n_{H}}$. For
 non-helical hydrodynamical turbulence  the Kolmogorov spectrum
 has $n_S=-11/3$.
 It has been speculated that in a magnetized medium   an
 Iroshnikov-Kraichnan spectrum with $n_S=-7/2$ might develop instead.
 The presence of hydrodynamical helicity makes the situation more
 complex. Two possibilities have been discussed. First,
with a
forward cascade
 of both energy and helicity
 (dominated by energy dissipation  on small scales) one has spectral
indices $n_S=-11/3$ and $n_H=-14/3$ (the helical Kolmogorov (HK)
spectrum), p.~243 of Ref.~\cite{L}.  Second, if helicity
 transfer and small-scale helicity dissipation dominate \cite{k73},
 $n_S=n_H=-13/3$ (the helicity transfer (HT) spectrum) \cite{MC96}.
 The HK spectrum has been observed in the
inertial range of weakly helical turbulence
 (i.e.,  $|P_H(k)| \ll P_S(k)$) \cite{BO97}.
 For strongly helical hydrodynamical turbulence
the characteristic
 length scale of helicity dissipation
 is larger than the
Kolmogorov energy dissipation length scale \cite{k73,DG01,helicity2}.
 Therefore
 the inertial range is taken to consist of two
sub-intervals, both with power-law spectra. For  smaller
$k$  the spectra are determined by helicity
 transfer and have
  $n_S=n_H=-13/3$, while for larger $k$
turbulence becomes non-helical and the more common HK spectrum is realized.
Since GW generation is mostly determined by the physics at small $k$
 \cite{kmk02,dolgov}, it is fair to only use the HT spectrum in this
 case also.

Based on these considerations  we model
$P_S(k)=S_0 k^{n_S}$ and $P_H(k)= A_0
k_S^{n_S-n_H}k^{n_H}$, where: (i) for the HK case $S_0=\pi^2C_k
{\bar\varepsilon }^{2/3}$ and $A_0=\pi^2C_k {\bar\delta}
/({\bar\varepsilon }^{1/3}k_S)$ \cite{DG01}, implying $A_0/S_0 =
{\bar \delta}/({\bar \varepsilon}k_S)$; and,   (ii) for the HT case  $S_0 =
\pi^2 C_s {\bar\delta }^{2/3}$ and $A_0 = \pi^2 C_a {\bar\delta
}^{2/3}$ \cite{MC96}. Here
 ${\bar\varepsilon}$ and  ${\bar\delta}$ are the energy and mean
 helicity
 dissipation rates per unit enthalpy, and $C_k$,
 $C_s$, and $C_a$ are  constants of order unity.

Given a  model of the turbulence, the turbulent source
two-point function is
\begin{eqnarray}
\langle\Pi^{\star}_{ij}({\mathbf k},t)\Pi_{lm}
({\mathbf k'},t+y)\rangle
~~~~~~~~~~~~~~~~~~~~~~~~~~~~~~~~~~~~
\nonumber\\
=(2\pi)^3 \delta^{(3)}({\bf k}-{\bf k'})\left[ {\mathcal
M}_{ijlm} f(k,y) + i{\mathcal A}_{ijlm} g(k,y) \right],
~\label{pi1}
\end{eqnarray}
where $ {\mathcal M}_{ijlm}$ and ${\mathcal A}_{ijlm}$ are defined below
Eq.~(\ref{gw1}). The functions $f(k, y)$ and $g(k,y)$  that
 describe the symmetric and helical
parts of the two-point function are
\begin{eqnarray}
f(k, y)&=&
\frac{(\rho+p)^2}{256\pi^6}
\!\int\!d^3p_1\!
\int\!d^3p_2 \delta^{(3)}({\bf k}-{\bf p_1}-{\bf p_2})
\nonumber \\
&&\times \left[(1+\gamma^2)(1+\beta^2)D_1^2(y)P_S(p_1)P_S(p_2)
+\right.
\nonumber \\
&& \left.
~~~~~~~~~~+4\gamma \beta D_2^2(y)
P_H(p_1)P_H(p_2)\right], \label{tensor-source-sym}
\\
g(k,y) &= &\frac{(\rho+p)^2 D_1(y)D_2(y)}{128\pi^6}\int\!d^3p_1\!\int\!d^3p_2
\,
\nonumber\\ &\times &
\delta^{(3)}({\bf k}-{\bf p_1}-{\bf p_2})
\left[ (1+\gamma^2)\beta P_S(p_1)P_H(p_2)
\right.
\nonumber \\
&&~~~~~~~~~~~~~~+
\left.
(1+\beta^2)\gamma P_H(p_1)P_S(p_2) \right],
\label{tensor-source-hel}
\end{eqnarray}
where $\gamma={\hat {\bf k}}\cdot{\hat {\bf p}_1}$ and $\beta={\hat
{\bf k}}\cdot{\hat {\bf p}_2}$.
 The
 helical source term $g(k, y)$ vanishes
for turbulence  without helicity.

To determine $H(k, t)$ and ${\mathcal H}(k, t)$ we
 solve Eq.~(\ref{h_evolution}) assuming that there is no GW
for times  $t<t_{\rm{in}}$, i.e., we choose  as initial conditions
$h_{ij}({\mathbf k}, t_{\rm{in}})=0={\dot h}_{ij}({\mathbf k},
t_{\rm{in}})$. To compute the induced GW power spectrum we use the
averaging technique described in detail in Sec. III.B
of Ref.~\cite{kmk02}. The main points are:
 (i) the
$\delta^{(3)}({\bf k}-{\bf k'})$ in Eqs.~(\ref{gw1}) and
(\ref{pi1})  ensure statistical isotropy of the GWs;
(ii) the statistical average can be approximated by either a time  or a
space average (this is justified for locally isotropic
turbulence
   see Sec. 21.2 of Ref.~\cite{my75});
(iii) we choose to time average since
the Green function for Eq.~(\ref{h_evolution}) and
the source term  $\Pi_{lm} ({\mathbf k},t)$ are time dependent.
  These approximations result
in Eq.~(32) of Ref.~\cite{kmk02},
\begin{eqnarray}
\langle h_{ij}^\star ({\mathbf k}, t_{\rm{fi}} ) h_{lm} ({\mathbf
k^\prime}, t_{\rm{fi}} )\rangle \simeq \frac{(16\pi G)^2 \tau
}{2kk^\prime} \int_{t_{\rm{in}}}^{t_{\rm{fi}}} dt~ \cos (kt)
\nonumber \\
\times \langle\Pi_{ij}^\star ({\mathbf k}, t_1) \Pi_{lm} ({\mathbf k^\prime},t_1+
t )\rangle. \label{gw-spectrum}
\end{eqnarray}
 The two-point  function on the r.h.s.~of the integral
is independent of $t_1$ and
$\langle h_{ij}^\star ({\mathbf k}, t_{\rm{fi}} ) h_{lm} ({\mathbf
k^\prime}, t_{\rm{fi}})\rangle $ is proportional to the
source duration time $\tau$,
 as  expected for locally isotropic turbulence,
see p.~358 of Ref.~\cite{my75}.

From Eqs.~(\ref{pi1})--(\ref{gw-spectrum}) we see that
 the symmetric
 $H(k,t)$ and helical
${\mathcal H}(k, t)$ parts  of the GW two-point function
 in Eq.~(\ref{gw1}) are
integrals over $y$ of $\cos (ky)
D_1^2(y)$, $\cos (ky)D_2^2(y)$, and $\cos (ky)D_1(y)D_2(y)$.
Taking into account that  both  $D_1(y)$ and $D_2(y)$
are positive monotonically-decreasing functions of $y$,
  $\int_{t_{\rm{in}}}^{t_{\rm{fi}}} dy \cos (ky) F_a(p,
y)F_b(|{\mathbf k}-{\mathbf p}|, y) \simeq P_a(p) P_b
(|{\mathbf k}-{\mathbf p}|)/(\sqrt{2} k)$
(where $a$ and $b$ can be $S$ or $H$).
Integrating over angles, we find at $t=t_{\rm{fi}}$,
\begin{eqnarray}
H(k) &\simeq & A \int\!dp_1~p_1\!\int\!dp_2~p_2
{\bar \Theta}\left[(1+\gamma_p^2)(1+\beta_p^2)P_S(p_1)
\right.
\nonumber \\
&\times & \left.P_S(p_2)+ 4 \gamma_p \beta_p P_H(p_1)P_H(p_2)\right],~~ \label{tensor-sym1_a}
\\
{\mathcal H}(k) &\simeq &2A\int\!
dp_1 ~p_1\!\int\!dp_2~p_2 {\bar \Theta}
\left[(1+\gamma_p^2)\beta_pP_S(p_1)
\right.
\nonumber \\
& \times& \left.P_H(p_2)+ (1+\beta_p^2)\gamma_pP_H(p_1)P_S(p_2) \right].~ \label{tensor-hel1_a}
\end{eqnarray}
Here $A= \alpha \tau/(4\pi^2 k^4)$ where $\alpha =
\sqrt{2} (p+\rho)^2 (8 \pi G)^2$,
$\gamma_p = (k^2+p_1^2-p_2^2)/(2kp_1)$, $\beta_p =
(k^2+p_2^2-p_1^2)/(2kp_2)$,  ${\bar \Theta} \equiv \theta(p_1+p_2-k)
\theta(p_1+k-p_2) \theta(p_2+k-p_1)$, and $\theta$ is the Heaviside
 step function which is zero (unity) for negative (positive) argument.

For power-law $P_{S}(k)\propto
k^{n_{S}}$ and  $P_{H}(k)\propto k^{n_{H}}$ the integrals in
Eqs.~(\ref{tensor-sym1_a}) and (\ref{tensor-hel1_a}) can be done
analytically, but the results are complicated and do not edify.  Instead
 we compute the degree of circular polarization, Eq.~(\ref{degree}), by
evaluating the integrals numerically for different
parameter values. Results are shown in  Fig. 1.
\begin{figure}
\includegraphics{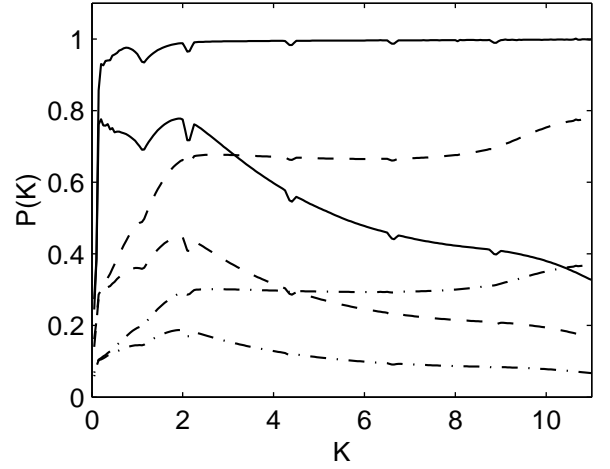}
\caption{GW polarization degree ${\mathcal P}(K, t_{\rm{fi}})$,
Eq.~(\ref{degree}), as a function of scaled wave number $K=k/k_S$
relative to the large-scale wave number $k_S$ on which energy is
pumped into the turbulence. This is evaluated at time
$t_{\rm{fi}}$, after the turbulence has switched off, and remains
unchanged to the present epoch. It has been computed for a damping
wave number $k_D=10k_S$. Three pairs of curves are shown. Solid
lines correspond to the amplitude ratio $A_0/S_0=1$ (maximally
helical turbulence \cite{helicity2}), dashed lines to
$A_0/S_0=0.5$, and dot-dashed lines are for $A_0/S_0 =0.2$. The
upper line in each pair corresponds to HT turbulence with
$n_S=n_H=-13/3$ \cite{MC96,DG01} and the lower line to HK
turbulence with $n_S=-11/3$ and $n_H=-14/3$ \cite{L}. Even for
helical turbulence with $A_0/S_0 \leq 0.5$,  for large wave
 numbers $k \sim k_D$,  $n_S=n_H=-13/3$ is unlikely so the large
$K$ part of the lower dashed and dot-dashed HT curves are
unrealistic. The large $k \sim k_D$ decay of the HK curves is a
consequence of vanishing helicity transfer at large $k$
\cite{k73}. }
\end{figure}

Figure 1 and other  numerical results show that for  maximal
helicity turbulence (when $A_0=S_0$) with equal spectral indices
$n_H = n_S<-3$, the polarization degree
${\mathcal P}(k) \simeq 1$ (upper solid line).
 For weaker helical turbulence (when $ A_0 <
S_0$) with $n_H \simeq  n_S<-3$,  ${\mathcal P(k)}
\rightarrow CA_0/S_0$, where $1<C(n_S,n_H)<2$ is a numerical factor that
depends on the spectral indices. For HT turbulence with
$n_S=n_H=-13/3$,  $C \approx 1.50$, while  for Iroshnikov-Kraichnan
MHD turbulence ($n_S=n_H=-7/2$), $C\approx1.39$. Excluding the edges of  the
inertial range  $k_S < k <k_D$,
an analysis of
Eqs.~(\ref{tensor-sym1_a}) and (\ref{tensor-hel1_a}) shows that the main
contribution to the integrals come from  areas with $p_1\sim
k_S,~p_2\sim k$ and $p_1\sim k,~p_2\sim k_S$.
In this case
(for arbitrary spectral indices $n_S$, $n_H<-3$)
 $H(k),~{\mathcal H(k)} \propto k^{n_S-3}
k_S^{n_S+3}$, so  this approximate analysis indicates that
  ${\mathcal P(k)}$ is
 not very sensitive to $k$.

In position space, we find
\begin{equation}
\langle  |h^\pm  ({\mathbf x},t_{\rm{fi}})|^2 \rangle \simeq
-\frac{ 9 \alpha \tau S_0^2 k_S^{n_S+3}}{64\pi^4 (n_S+3)}
\int_{k_S}^{k_D}dk (1 \pm {\mathcal P}(k))k^{n_S-1}.
\label{H43}
\end{equation}
GW amplitudes  are  conventionally expressed as
 $
\langle  h^{ij}  ({\mathbf x},t_{\rm{fi}})
 h^{ij}  ({\mathbf x},t_{\rm{fi}}) \rangle
  = 2 \int_0^\infty d{\rm{ln}}f [|h^{+}(f)|^2 + |h^{-}(f)|^2],
$ see Eq.~(11) of Ref.~\cite{m00}, where
the frequency $f$ of a GW generated
  by an eddy of length $L$ is
$f=\tau_L^{-1}$ where $\tau_L$ is the eddy turnover time
\cite{kos,kmk02,dolgov}.
This is  inversely  proportional to
the cosmological scale factor, so the
frequency ${\bar f}$ today and $f$  when the temperature was
$T_{\rm{in}}=100~T_{100}$ GeV are related by
${\bar f} = 1.65 \times 10^{-5}  T_{100} g_{100}^{1/6}
f/H_{\rm{in}}$ Hz  \cite{m00},
 where $g_{\rm{in}} = 100 g_{100}$ is the number of
 relativistic degrees of freedom at
$t_{\rm{in}}$. Since we truncate turbulence power for $L>L_S$,
the GW spectrum is non-zero only for
${\bar f}>{\bar f}_S$, where
\begin{eqnarray}
{\bar f}_S &=&  1.9 \times 10^{-6}
\sqrt{\frac{n_S+5}{|n_S+3|}}
\left(\frac{{\bar\varepsilon}}{ \nu}\right)^{{1}/{2}}
\left(\frac{L_D}{L_S}\right)^{(n_S+5)/{2}}
\nonumber \\ && \times
T_{100}~g_{100}^{{1}/{6}}~H_{\rm{in}}^{-1}~{\mbox{Hz}}
\label{fs}
\end{eqnarray}
is the frequency now that corresponds to the stirring length $L_S$.
Here we use the fact that for locally isotropic turbulence the energy
dissipation rate  ${\bar\varepsilon} =
2\nu \int_{k_S}^{k_D} dk~k^4 P_S(k)/\pi^2$
(where $\nu$ is the plasma viscosity, p.~483 of Ref.~\cite{my75}),
 is equal to the source power input, i.e.,
${\bar\varepsilon} =
3\kappa\rho_{\rm{vac}}/(4\rho\tau)$ and $\kappa$ is the phase transition
efficiency. For HK turbulence with $n_S=-11/3$, Eq.~(\ref{fs}) is Eq.~(53)
of Ref.~\cite{kmk02}.

Using Eqs.~(\ref{H43}) and (\ref{fs}),
 and neglecting the weak $k$-dependence
of the GW polarization degree ${\mathcal P}$,  we find
 that $h^\pm ({\bar f}) \propto {\bar f}^{-11/4}$ for the HK case
 \cite{kmk02,dolgov}, while for HT turbulence  $h^\pm ({\bar f})\propto
 {\bar f}^{-13/2}$. We expect such  a steeper dependence on
frequency for helicity induced
GWs, since the helicity transfer rate is more important on
larger scales \cite{k73}.  In both cases the amplitude of the
GW spectrum peaks at the stirring frequency ${\bar f}_S$.

We close with a brief examination of the prospect of detecting such circularly polarized GWs.
The GW energy density parameter for frequency
${\bar f}$,
$\Omega_{\rm{GW}} ({\bar f})$ is given by (see Eq.~(7) of Ref~
\cite{m00})
$\Omega_{\rm{GW}}({\bar f}) h^2 = 5.9 \times 10^{35}
(|h^+({\bar f})|^2 + |h^-({\bar f})|^2)({\bar f}/\mbox{Hz})^2$, where
$h$ is the Hubble constant in  units of $100$ km~sec$^{-1}$ Mpc$^{-1}$. In our case,
\begin{eqnarray}
&&\Omega_{\rm{GW}}({\bar f}) h^2 \simeq   1.05 \times 10^{-11}~g_{100}^{-{1}/{3}}
\left(\frac{L_S^2}{\tau H_{\rm{in}}^{-1}}\cdot \frac{n_S+5}{|n_S+3|}\right)^2
\nonumber \\ &&\times
\left(\frac{L_D}{L_S}\right)^{3(n_S+5)}
\left(\frac{3\kappa \rho_{\rm{vac}} L_S}{4\nu \rho}\right)^3
\left(\frac{\bar{f}}{\bar{f}_S}\right)^{2(2n_S+5)/(n_S+5)}\!\!\!\!\!\!\!\!\!\!\!\!\!\!\!\!\!\!\!\!\!\!\!\!\!\!\!\!\!\!\!\!\!\!\!.
\label{energy}
\end{eqnarray}

The stirring frequency ${\bar f}_S$ and the GW spectrum are very sensitive
to phase transition properties.
If the phase transition is strongly first order, for the HK case
 ${\bar f}_S  \simeq 5 \times 10^{-3}$Hz
near the Laser Interferometer Space Antenna (LISA)  frequency
range, but the amplitude $\Omega_{\rm{GW}}({\bar f}) h^2 \simeq
10^{-11}$ \cite{kmk02,dolgov} is below  LISA sensitivity on this
frequency \cite{nic,tasi}. An additional limit of detectability is
imposed by
 the dominating
 stochastic
GWs signal from white
dwarf binaries
\cite{white-dwarf}.
 Thus it
 is unlikely that the
 GWs discussed here will be detected by currently planed GW detectors,
 but  future detector configurations  \cite{GWdetection}
 may well be able to.

\acknowledgments
We thank A. Kosowsky for fruitful discussions
and suggestions. We also acknowledge helpful comments from
A.~Brandenburg, A. Dolgov,
R. Durrer, D. Grasso, K. Jedamzik, T. Vachaspati,
and L. Weaver.
 We acknowledge support
from CRDF-GRDF grants 3315 and 3316, NSF CAREER grant AST-9875031, and DOE EPSCoR grant DE-FG02-00ER45824.


\end{document}